\documentclass[letterpaper, 10 pt, conference]{ieeeconf}  
\IEEEoverridecommandlockouts                              
\title{\LARGE \bf
Sieving out Unnecessary Constraints in Scenario Optimization \\ with an Application to Power Systems}

\author{Miguel Picallo and Florian D{\"o}rfler
\thanks{Funding by the Swiss Federal Office of Energy through the project “Renewable Management and Real-Time Control Platform (ReMaP)” (SI/501810-01) and the ETH Foundation is gratefully acknowledged.}
\thanks{M. Picallo and F. D{\"o}rfler are with the Automatic Control Laboratory at ETH Z{\"u}rich, Switzerland.
{\tt\small \{miguelp,doerfler\}@ethz.ch}}%
\thanks{This version includes some corrections to the published version \cite{picallo2019scerem}.}
}

\usepackage{graphicx}      

\usepackage{commath}
\usepackage{mathtools}
\usepackage{steinmetz}		
\usepackage{supertabular}	
\usepackage{tabularx}		
\usepackage{algorithm}		
\usepackage{algorithmic}		
\usepackage{amsmath,amssymb}
\newtheorem{rem}{Remark}
\newtheorem{defi}{Definition}

\newtheorem{thm}{Theorem}
\newtheorem{ass}{Assumption}
\usepackage{subfig}
\usepackage{color}
\usepackage{soul}
\usepackage{url} 
\usepackage{afterpage}

\renewcommand{\r}{\textcolor{black}}

\begin{document}


\maketitle

\begin{abstract}
Many optimization problems incorporate uncertainty affecting their parameters and thus their objective functions and constraints. As an example, in chance-constrained optimization the constraints need to be satisfied with a certain probability. To solve these problems, scenario optimization is a well established methodology that ensures feasibility of the solution by enforcing it to satisfy a given number of samples of the constraints. The main theoretical results in scenario optimization provide the methods to determine the necessary number of samples, or to compute the risk based on the number of so-called support constraints. In this paper, we propose a methodology to remove constraints after observing the number of support constraints and the consequent risk. Additionally, we show the effectiveness of the approach with an illustrative example and an application to power distribution grid management when solving the optimal power flow problem. In this problem, uncertainty in the loads converts the admissible voltage limits into chance-constraints. 
\end{abstract}

\IEEEpeerreviewmaketitle

\section{Introduction}

Many real-world optimization problems are full of uncertainties, but this is ignored in many cases. However, if this uncertainty is not taken into account, it may lead to suboptimal solutions, or what is even worse, solutions that do not satisfy the constraints of the problem. Therefore, it is essential to represent these uncertainties with random variables and stochastic constraints.

There are mainly two methodologies to deal with stochastic constraints: robust optimization and chance-constrained optimization. Robust optimization \cite{bental1998robustopt} aims at ensuring feasibility for any realization of the random variables in the stochastic constraints, and thus prevents the worst-case scenarios. However, robust optimization solutions may be too conservative in terms of performance, since improbable extreme values of the random variables may shrink the feasible region too much. In the worst case, this may even lead to problems with an empty feasible space.

On the other hand, in chance-constrained optimization the constraints do not need to be satisfied for every realization, but only satisfied with a given probability guarantee, like $90\%$, $95\%$ or $99\%$. These chance-constrained optimization problems are typically solved by turning the stochastic constraints into deterministic ones. Sometimes this can be done exactly, for example, in the case of Gaussian uncertainty with linear constraints; in other cases, a conservative approach can be used to ensure feasibility \cite{nemirovski2006approxchanceconstr}. However, in many cases the problem may not be tractable due to various reasons, like not knowing the probability distribution of the random variables.

The scenario approach \cite{Calafiore2005support, calafiore2006scenario, campi2008exact, campi2011sampling, campi2018wait} is a data-driven optimization that aims at solving these intractable chance-constrained optimization problems by using only samples of scenarios, i.e., realizations of the random variables. These scenarios are used to subtitute the chance-constraint by a deterministic constraint, one for every scenario. In \cite{campi2008exact} a method is derived to determine the necessary number of scenarios to ensure a given probability guarantee. However, this method may result in too conservative solutions, that have an expected risk far below the risk threshold of the original chance-constrained problem. Therefore, in \cite{campi2011sampling} this method is improved by allowing to violate some scenarios. 
Support constraints are defined as the scenarios whose removal results in an improvement of the objective function. In \cite{campi2008exact} it is proven that the scenario approach produces exact bounds, when the solution of the problem has as many support constraints as decision variables. This is called a fully-supported solution. When this is not the case, a better bound on the risk can be computed using the method in \cite{campi2018wait} as a function of the number of support constraints.

The natural question arising is then the following: when a not fully-supported solution is observed, can the optimal value of the objective function be improved, while still keeping the risk below its limit? The contribution of this paper consist in proposing an extension of \cite{campi2018wait}, to sieve out scenarios like in \cite{campi2011sampling}, when a not fully-supported solution is found. An illustrative example is presented to show the effectiveness of the approach. Moreover, a power system application is presented. In this application, our scenario approach allows to find a feasible solution, that is closer to the true optimum than the standard scenario approach \cite{campi2008exact}.

In the context of power systems, multiple uncertainties may affect the modelling of the optimization problems. An example is the optimal power flow, an optimization problem used to determine the set-points of controllable elements in an electrical grid. Uncertainties in loads and generation availability need to be taken into account in order to provide solutions satisfying the constraints. In \cite{summers2015stochopf, dall2017chance, picallo2018stochOPF} this is achieved assuming a known probability distribution of these uncertainties. In \cite{bolognani2016fastPSCC, bolognani2017fast, ming2019scedemresp, modarresi2018scerisks, roald201566} this assumption is dropped, and the scenario approach is used as an appropriate tool to solve this problem.

The rest of the paper is structured as follows: Section \ref{sec:prevwork} describes some relevant definitions and results established in the context of scenario optimization. Section \ref{sec:remsupsce} presents the main contribution of the paper: the method to remove constraints after observing the number of support constraints, with an example to illustrate its effectiveness. Section \ref{sec:powersys} presents the power system application. Finally, Section \ref{sec:conclusions} draws some conclusions and proposes future work. 

\section{Definitions and results in scenario approach}\label{sec:prevwork}
Here we present some relevant definitions and results of the scenario approach \cite{calafiore2006scenario, campi2008exact, campi2011sampling, campi2018wait} for chance-constrained optimization problems, which will be necessary to present our approach.

\begin{defi}[chance-constrained optimization] 
A chance-constrained problem $CCP$ can be defined as
\begin{equation}\label{eq:CCP}
CCP: x^* = \arg \hspace{-0.3cm} \min_{x \in \mathcal{X} \subseteq \mathbb{R}^d} \hspace{-0.1cm} c^T x \text{ s.t. } P( \delta \in \Delta : x \in \mathcal{X}_{\delta} ) \geq 1 - \epsilon,
\end{equation}
where $d$ is the size of the decision variables $x$, $\delta$ is a random variable with support $\Delta$ included in a measurable $\sigma$-algebra, $\mathcal{X}_{\delta}$ denotes the set of feasible solutions, which depends \r{on multiple uncertainties represented through the vector of random variables} $\delta$, and $\epsilon$ is the risk of violating the constraints. 
\end{defi}

As in \cite{campi2008exact}, we will consider throughout this paper the case where $\mathcal{X}$ and $\mathcal{X}_{\delta}$ are closed convex sets; and we will assume that the feasibility domain of problem \eqref{eq:CCP} has a nonempty interior, and that a solution always exists and is unique.

The problem \eqref{eq:CCP} may be hard to solve depending on the shape of the feasible space $\mathcal{X}_{\delta}$, and the probability distribution of $\delta$. This probability may even not be known. Therefore, the scenario approach provides a technique to solve the problem using only samples of $\delta$ or equivalently $\mathcal{X}_{\delta}$.

\subsection{Exact feasibility in the scenario approach \cite{campi2008exact}}\label{sec:basicsce}

The scenario approach is based on enforcing the constraints for a given number of samples of $\delta$:
\begin{defi}[scenario approach] 
The scenario approach problem $SP_N$ with $N$ samples: $\delta^{(i)}, \; i \in \mathcal{N}=\{1,\dots,N\}$, can be defined as
\begin{equation}\label{eq:sceproblem}
SP_N: x_N^* = \arg\hspace{-0.3cm}\min_{x \in \mathcal{X} \subseteq \mathbb{R}^d} c^T x \text{ s.t. } x \in \bigcap_{i\in \mathcal{N}} \mathcal{X}_{\delta^{(i)}},
\end{equation}
where $\mathcal{X}_{\delta^{(i)}}$ is the feasible space given that the random variable takes value $\delta^{(i)}$.
\end{defi}

The notion of violation probability and support constraints are key to understand the scenario approach. They can be defined as follows:
\begin{defi}[violation probability] 
The violation probability of a solution $x$ is $V(x) = P(\delta \in \Delta : x \notin \mathcal{X}_{\delta})$. 
\end{defi}
The solution $x_N^*(\delta^N)$ of \eqref{eq:sceproblem} and its violation probability $V(x_N^*(\delta^N))$ are random variables, since they depend on the set of scenarios sampled $\delta^N=(\delta^{(1)},\dots,\delta^{(N)}) \in \Delta^N$. In an abuse of notation, we denote by $P(\delta^N \in \Delta^N : V(x_N^*(\delta^N))>\epsilon) = P(V(x_N^*)> \epsilon)$ the probability (with respect to the samples $\delta^N$) of the violation probability.
\begin{defi}[support constraint \cite{calafiore2006scenario}] 
Consider the problem without scenario $k$:
\begin{equation*}
SP_{N,k}: x_{N,k}^*= \arg\hspace{-0.3cm}\min_{x \in \mathcal{X} \subseteq \mathbb{R}^d} c^T x \text{ s.t. } x \in \hspace{-0.2cm} \bigcap_{i\in \mathcal{N}\setminus\{k\}} \hspace{-0.2cm} \mathcal{X}_{\delta^{(i)}},
\end{equation*}
scenario $k$ is a support constraint if $c^T x_{N,k}^* < c^T x_N^*$. 
\end{defi}

The set of indices of the support constraints for the solution $x_N^*$ is denoted as $S_N^*$. There can be at most $d$ support constraints \cite{Calafiore2005support}, $|S_N^*| \leq d$, where $|\cdot|$ denotes the set cardinality. If $|S_N^*| = d$ then the problem is said to be fully-supported. These definitions allow to introduce the following theorem:

\begin{thm}
\cite{campi2008exact} Given a number of samples $N$, the probability of the violation probability $V(x_N^*)$ exceeding a risk value $\epsilon$ can be bounded by the term $\beta$ as
\begin{equation}\label{eq:probfeas}
P(V(x_N^*)> \epsilon) \leq \beta := \sum_{i=0}^{d-1} \binom{N}{i} \epsilon^i (1-\epsilon)^{(N-i)},
\end{equation}
where equality holds if the problem is fully-supported.
\end{thm}

Then, fixing the risk parameter $\epsilon$ and $\beta$, this theorem allows to find the right number of scenarios $N$ to limit the probability of $V(x_N^*)$ being above $\epsilon$. 

\subsection{Scenario approach discarding constraints \cite{campi2011sampling}} \label{sec:remsce}

In \cite{campi2011sampling} a method is proposed to trade feasibility for performance by allowing a number of constraints to be violated. This allows to get rid of outlier scenarios that would shrink the feasible space too much.

\begin{defi}[scenario approach discarding constraint]
The scenario approach $SP_{N,R}^{\mathcal{A}}$ with $N$ samples, where $R$ are selected to be violated using some removal algorithm $\mathcal{A}(\cdot)$ such that $\abs{\mathcal{A}(\mathcal{N})} = R$, can be defined as:
\begin{equation}\label{eq:sceremprob}\arraycolsep=1pt\begin{array}{rl}
SP_{N,R}^{\mathcal{A}}: x_{N,R}^* = & \arg\min_{x \in \mathcal{X} \subseteq \mathbb{R}^d} c^T x \\[0.2cm] 
\text{ s.t. } & x \in \bigcap_{i \in \mathcal{N} \setminus \mathcal{A}(\mathcal{N})} \mathcal{X}_{\delta^{(i)}}.
\end{array}
\end{equation}
where any removal algorithm $\mathcal{A}$ could be valid. The only requirement for the following theorem, is that the constraints removed by algorithm $\mathcal{A}$ are almost surely violated. This can be achieved be testing if removed constraints are effectively violated, and if not, remove other constraints. 
\end{defi}

\begin{thm}
\cite{campi2011sampling} If the solution $x_{N,R}^*$ almost surely (with respect to the set of samples $\delta^N$) violates the $R$ constraints removed by $\mathcal{A}$ out of the total number of constraints $N$; then the probability of the violation probability $V(x_N^*)$ exceeding a risk value $\epsilon$ can be bounded by $\beta$ as
\begin{equation}\label{eq:probfeasrem}
P(V(x_{N,R}^*) \hspace{-0.06cm} > \hspace{-0.05cm} \epsilon) \hspace{-0.06cm} \leq \hspace{-0.05cm} \beta \hspace{-0.05cm} := \hspace{-0.06cm} \binom{ \hspace{-0.05cm} R+d-1 \hspace{-0.05cm}}{R} \hspace{-0.2cm} \sum_{i=0}^{R+d-1} \hspace{-0.2cm} \binom{ \hspace{-0.05cm} N}{i} \epsilon^i (1-\epsilon)^{( \hspace{-0.03cm} N-i)} \hspace{-0.1cm}.
\end{equation}
\end{thm}

Note that for the same $N$ the right-hand term of \eqref{eq:probfeasrem} is larger that the one in \eqref{eq:probfeas}. Therefore, to achieve the same bound on $P(V(x_{N,R}^*) > \epsilon)$, the approach with removals requires a larger number of samples $N$. The approach in \eqref{eq:sceremprob},\eqref{eq:probfeasrem} remedies a drawback in the basic scenario approach in \eqref{eq:sceproblem},\eqref{eq:probfeas}: the probability density of $V(x_N^*)$ may be concentrated close to $0$ and so the expected value, $\mathbb{E}[V(x_N^*)] \ll \epsilon$. On the other hand, for the optimal solution $x^*$ of \eqref{eq:CCP} we would expect to have $V(x^*)$ close to $\epsilon$, if that helps to improve the performance. This means that the basic scenario approach may be too conservative. The approach in \eqref{eq:sceremprob},\eqref{eq:probfeasrem} can produce less conservative solutions with  $\mathbb{E}[V(x_{N,R}^*)]$ closer to $\epsilon$ \cite[Appendix A]{campi2011sampling}.

\subsection{Risk given support constraints \cite{campi2018wait}}\label{sec:supsce}

In \cite{campi2018wait} the authors analyze the case of not fully-supported problems. Instead of having a fixed risk $\epsilon$, it is defined as function of the number of support constraints $|S_N^*|$ of the solution $x_N^*$ of \eqref{eq:sceproblem}: $\epsilon(|S_N^*|)$. Throughout the rest of the paper, we will make the following assumtion as in \cite{campi2018wait}:

\begin{ass}[Non-degeneracy]\label{ass:nondeg}
For every $N$ and $R$, with probability $1$ (with respect to $\delta^N$) the solution of \eqref{eq:sceproblem} and \eqref{eq:sceremprob} with all the constraints, except removed constraints in \eqref{eq:sceremprob}, coincides with the solution where only the support constraints are kept.
\end{ass}

With Assumption \ref{ass:nondeg} in place we have the following result:
\begin{thm}\cite{campi2018wait}
Given $\beta \in (0,1)$, for any $k =1,\dots,d$ indicating the number of support constraints, the polynomial equation in $\epsilon$ given by
\begin{equation}\label{eq:probfeassup}
0=\frac{\beta}{N+1} \sum_{m=k}^N \binom{N}{k}(1-\epsilon)^{m-k}-\binom{N}{k}(1-\epsilon)^{N-k},
\end{equation} 
has exactly one solution $\epsilon(k) \in (0,1)$, and we have
\begin{equation*}
P(V(x_{N}^*) > \epsilon(|S_N^*|) ) \leq \beta.
\end{equation*}
\end{thm}

It can be observed in \cite[Figure 1]{campi2018wait} and in Figure \ref{fig:risksuprem} that the number of samples $N$ computed for a fully-supported problem, produces too conservative solutions in problems with a low number $k$ of support constraints, because $\epsilon(k) < \epsilon$ for a low $k$. Therefore, the approach in \eqref{eq:probfeassup} gives a better bound on the risk than \eqref{eq:probfeas}.

\section{Scenario approach discarding constraint given support constraints}\label{sec:remsupsce}

On the one hand, the method in Section \ref{sec:supsce} allows to determine a better bound on the risk $\epsilon$ as a function of the number of support constraints $k$: $\epsilon(k)$. However, it does not provide a method to get a less conservative solution if the number of support constraints is low. On the other hand, the method in Section \ref{sec:remsce} allows to obtain potentially less conservative solutions, but it does not take into account the number of support constraint to get the bound on the risk. Therefore, in this section we present a method to react by removing constraints, when having a not fully-supported problem, and thus an $\epsilon(k)$ lower than the $\epsilon$ intended. This method is in essence a combination of the methods in presented in Sections \ref{sec:remsce} and \ref{sec:supsce} with quantitative guarantees. 

Consider the problem $SP_{\tilde{N},R}^{\mathcal{A}}$, same as in \eqref{eq:sceremprob}, but with a total number of samples $\tilde{N}=N+R$ before removing $R$ samples. Here $\tilde{\mathcal{N}}$ denotes the set of all samples; $x_{\tilde{N},R}^*$ is the solution of the problem; and $S_{\tilde{N},R}^*$ the set of indices of its support constraints. Then, again with Assumption \ref{ass:nondeg}, we have the following theorem, which is the main result and contribution of this paper:
\begin{thm}\label{thm:remsup}
Given $\beta \in (0,1)$, for any $k =1,\dots,d$ indicating the number of support constraints, and $R$ the number of removed constraints, the equation in $\epsilon$
\begin{equation}\label{eq:probfeassuprem}
0 \hspace{-0.05cm} = \hspace{-0.05cm} \frac{\beta}{N+1} \hspace{-0.1cm} \sum_{m=k}^N \hspace{-0.15cm} \binom{N}{k} \hspace{-0.05cm} (1-\epsilon)^{m-k}-\binom{N+R}{R} \hspace{-0.1cm} \binom{N}{k} \hspace{-0.05cm} (1-\epsilon)^{N-k},
\end{equation} 
has exactly one solution $\epsilon(k,R) \in (0,1)$. If the solution $x_{\tilde{N},R}^*$ almost surely (with respect to the sample $\delta^{\tilde{N}}$) violates $R$ constraints; then the probability of the violation probability $V(x_{\tilde{N},R}^*)$ exceeding a risk value $\epsilon$ can be bounded by $\beta$:
\begin{equation*}
P(V(x_{\tilde{N},R}^*) > \epsilon(|S_{\tilde{N},R}^*|,R) ) \leq \beta.
\end{equation*}
See Appendix \ref{sec:appproofthm} for the proof.
\end{thm}

\begin{figure}[!t]
\centering
\includegraphics[width=8.5cm]{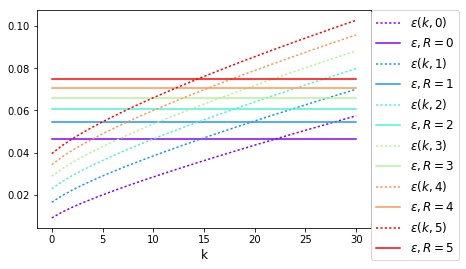}   
\caption{Values of $\epsilon(k,R)$ for a different number $R$ of violated constraints, with $N=1000, d=30, \beta = 10^{-3}$. Continuous horizontal lines of $\epsilon$ for different $R$ represent the bounds produced by methods described in Section \ref{sec:prevwork}, in \eqref{eq:probfeas} and \eqref{eq:probfeasrem}; while dotted lines represent those of Section \ref{sec:remsupsce} in \eqref{eq:probfeassuprem}.} 
\label{fig:risksuprem}
\end{figure}

Note that the polynomial expression in \eqref{eq:probfeassuprem} differs from the one in \eqref{eq:probfeassup} by the factor $\binom{N+R}{R} \geq 1$. As a consequence, for an $\epsilon(k)$ satisfying \eqref{eq:probfeassup}, we will have $\epsilon(k,R) \geq \epsilon(k)$  to satisfy \eqref{eq:probfeassuprem}. So the risk bound increases as constraints are removed. Then, it is possible to get solutions closer to the desired risk, when observing a low number of support constraints. In Figure \ref{fig:risksuprem}, it can be observed that the risk of a solution with a low number of support constraints and $R$ removed constraints, can be lower than the risk of fully-supported solutions satisfying all constraints.

\begin{rem}[Implementation]
For any scenario optimization problem, once $N$ is fixed using an approach like \eqref{eq:probfeas} or \eqref{eq:probfeasrem}, the solutions $\epsilon(k,R)$ of \eqref{eq:probfeassuprem} can be computed for all $k$ and $R$. These solutions can be stored in a look-up table, so that they can be quickly accessed for any number $k$ of support constraints observed and $R$ constraints removed.
\end{rem}

\subsection{Illustrative example}\label{sec:example}

\begin{figure}[!t]
\centering
\includegraphics[width=3.5cm]{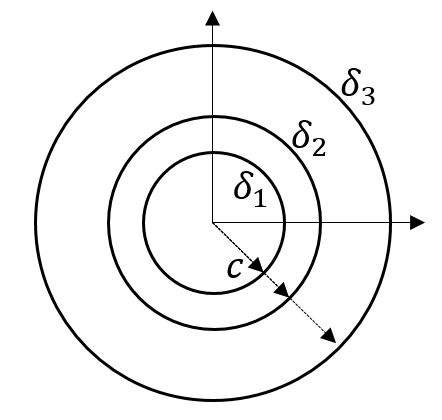}  
\caption{Example with random radiuses $\delta_i$.}
\label{fig:example}
\end{figure}

In this section we show the effectiveness of the approach with an illustrative example. Consider the following chance-constrained optimization problem
\begin{equation}\label{eq:example}
\min_{x \in \mathbb{R}^{d}} c^Tx 
\mbox{ s.t. } P(\delta : ||x||_2 \leq \delta) \geq 1-\epsilon,
\end{equation}
where $\delta$ is a random variable, and $\epsilon$ is the risk threshold. This example corresponds to minimizing a linear function subject to the constraint that the variables are within a sphere of random radius, see Figure \ref{fig:example}. Therefore, when applying scenario optimization, the sphere with the smallest radius will be the single support constraint, and thus the number of support constraints will be $k=1$ almost surely. 

In Table \ref{tab:example} we compare the results for two probability distributions using the approaches described in \eqref{eq:probfeas}, \eqref{eq:probfeasrem} and \eqref{eq:probfeassuprem}, in Sections \ref{sec:basicsce}, \ref{sec:remsce} and \ref{sec:remsupsce} respectively. In this example, since the probability distribution of $\delta$ is known we can determine the exact maximum value of $\delta_\epsilon$ such that $||x||_2 \leq \delta_{\epsilon} \implies P(||x||_2  \leq \delta) \geq 1-\epsilon$.  

First, it can be observed that the method described in Section \ref{sec:remsupsce} performs significantly better than the ones from Sections \ref{sec:basicsce} and \ref{sec:remsce}. However, it still relatively far away from the maximum value $\delta_{\epsilon}$. 
Note that for \ref{sec:basicsce}, negative values indicate that the problem would be infeasible. When increasing the dimension $d$ of the problem, while keeping the number of support constraints $k$ constant, from Table a) to b), the difference between methods further increases. 

\begin{table}[!t]

\centering

\subfloat[$\epsilon = 0.05, d = 30$]{
\begin{tabular}{|l|l|l|l|l|}
\hline
case & \hspace{-0.4cm} \begin{tabular}{l} exact \\ maximum  \\ value \end{tabular} \hspace{-0.4cm}  & 
\hspace{-0.4cm} \begin{tabular}{l} \ref{sec:basicsce} \eqref{eq:probfeas} \\ $(N=923)$ \end{tabular} \hspace{-0.4cm} & 
\hspace{-0.4cm} \begin{tabular}{l} \ref{sec:remsce} \eqref{eq:probfeasrem} \\ $(N=1535,$ \\ $r=5)$ \end{tabular} \hspace{-0.4cm} &  
\hspace{-0.4cm} \begin{tabular}{l} \ref{sec:remsupsce} \eqref{eq:probfeassuprem} \\ $(N=923,$ \\ $r=5)$ \end{tabular} \hspace{-0.4cm}  \\ \hline \hline
$\delta \sim \mathcal{N}(3,1)$ & $1.35$ & $-0.21$ & $0.31$ & $0.49$ \\ \hline
$\delta \sim \mathcal{U}(0,1)$ & $0.05$ & $0.001$ & $0.004$ & $0.007$ \\ \hline
\end{tabular}
}\label{tab:res530}


\hfill

\subfloat[$\epsilon = 0.05, d = 100$]{
\begin{tabular}{|l|l|l|l|l|}
\hline
case & \hspace{-0.4cm} \begin{tabular}{l} exact \\ maximum  \\ value \end{tabular} \hspace{-0.4cm}  & 
\hspace{-0.4cm} \begin{tabular}{l} \ref{sec:basicsce} \eqref{eq:probfeas} \\ $(N=2230)$ \end{tabular} \hspace{-0.4cm} & 
\hspace{-0.4cm} \begin{tabular}{l} \ref{sec:remsce} \eqref{eq:probfeasrem} \\ $(N=4920,$ \\ $r=17)$ \end{tabular} \hspace{-0.4cm} &  
\hspace{-0.4cm} \begin{tabular}{l} \ref{sec:remsupsce} \eqref{eq:probfeassuprem} \\ $(N=2230,$ \\ $r=17)$ \end{tabular} \hspace{-0.4cm}  \\ 
\hline \hline
$\delta \sim \mathcal{N}(3,1)$ & $1.35$ & $-0.46$ & $0.31$ & $0.58$ \\ \hline
$\delta \sim \mathcal{U}(0,1)$ & $0.05$ & $0.0004$ & $0.004$ & $0.008$ \\ \hline
\end{tabular}
}\label{tab:res5100}


\caption{Values of $\delta_{\epsilon}$ for bound $\beta =10^{-3}$ using the different scenario approaches described in \eqref{eq:probfeas}, \eqref{eq:probfeasrem}, \eqref{eq:probfeassuprem}. Two probability distributions are used: the normal distribution with mean $3$ and standard deviation $1$: $\mathcal{N}(3,1)$; and the uniform distribution in $(0,1)$: $\mathcal{U}(0,1)$. For the scenario approaches, the values correspond to the mean of 10000 sets of samples with $N$ samples each.}\label{tab:example}
\end{table}

\section{Power System Application}\label{sec:powersys}
In power systems, the optimal power flow consists of solving an optimization problem to determine the set-points of controllable elements, like the power injection of distributed energy sources. Additionally, the grid constraints like voltage limits need to be satisfied, but this may be difficult, because some parameters in this optimization problem may be uncertain. An example could be distribution grid management, where typically only a few measurements are available \cite{picallo2017twostepSE}, and thus the actual value of bus loads of the grid may be unknown. Here we combine the models and settings in \cite{picallo2018stochOPF} and \cite{bolognani2017fast} to build a simulation framework and apply our scenario approach \eqref{eq:probfeassuprem} to power systems.

First, we consider a linear approximation of the voltage magnitudes corresponding to the \textit{Linear Coupled power flow model} \cite[Section 5]{bolognani2016existence}
\begin{equation}\label{eq:PFlin}
\abs{V} = \abs{V_0} + \text{diag(\abs{V_0})}^{-1}(Z_p(P_G + P_L) + Z_q(Q_G+Q_L)),
\end{equation}
where $P_L,Q_L$ are the active and reactive loads respectively, which we consider to be uncertain parameters; $P_G,Q_G$ are the active and reactive generation, which we optimize; $Z_p,Z_q$ are the known impedance matrices of the grid; $\abs{V}$ is the vector of voltage magnitudes for all nodes; and $\abs{V_0}$ is the known vector of voltage magnitudes under no load and generation ($P_L,Q_L,P_G,Q_G=0$). Since $P_L,Q_L$ are unknown, so will be $\abs{V}$. 

When changing the values of the current operating point $P_G,Q_G$ by increments $\Delta P_G,\Delta Q_G$
, we induce a change in the voltage magnitudes:
\begin{equation}\label{eq:PFlinincr}
\abs{V}_\text{new} = \abs{V} + \text{diag(\abs{V_0})}^{-1}(Z_p \Delta P_G + Z_q\Delta Q_G).
\end{equation}
The advantage of the expression \eqref{eq:PFlinincr} over \eqref{eq:PFlin} is that the uncertain parameters $P_L,Q_L$ no longer appear in \eqref{eq:PFlinincr}. The single uncertain paremeter is the vector $\abs{V}$, since during the optimization process we have $\Delta P_L,\Delta Q_L=0$ \cite{picallo2018stochOPF}. Therefore, we only need to generate samples of $\abs{V}$ for the scenario optimization, see later in \eqref{eq:PFsample}. Moreover, since $\abs{V}$ appears as an additive term in \eqref{eq:PFlinincr}, this simplifies the process of identifying support scenarios and scenarios to remove, as we will see later in \eqref{eq:PFsampleelem}.

An objective for the optimal power flow could be to maximize the injection of distributed energy sources, and thus minimize the energy required from the grid, while satisfying the voltage limits $V_{\min}, V_{\max}$:
\begin{equation}\label{eq:OPF}\arraycolsep=1pt\begin{array}{l}
\max \sum_i \Delta P_{G,i} + \Delta Q_{G,i} \\[0.2cm]
\text{s.t. } V_{\min}\textbf{1} \leq \abs{V}_\text{new}(\Delta P_G,\Delta P_G,\abs{V}) \leq V_{\max}\textbf{1},
\end{array}
\end{equation}
where $\textbf{1}$ is a vector of $1$, and $\abs{V}_\text{new}(\Delta P_G,\Delta P_G,\abs{V})$ denotes $\abs{V}_\text{new}$ as a function of $\Delta P_G,\Delta P_G,\abs{V}$, by enforcing constraint \eqref{eq:PFlinincr}. Since $\abs{V}$ in \eqref{eq:PFlinincr} is an uncertain parameter, the constraints in \eqref{eq:OPF} will be stochastic. We would like to satisfy these stochastic constraints with a certain probability guarantee, for example $95\%$:
\begin{equation}\label{eq:OPFstoch}\arraycolsep=1pt\begin{array}{l}
\max \sum_i \Delta P_{G,i} + \Delta Q_{G,i} \\
\text{s.t. } \\
P(\abs{V} \hspace{-0.04cm} : 
V_{\min}\textbf{1} \hspace{-0.05cm} \leq 
\abs{V}_\text{new}(\Delta P_G,\Delta P_G,\abs{V}) 
\leq \hspace{-0.05cm} V_{\max}\textbf{1}) 
\geq 95\% .
\end{array}
\end{equation}

Historical data or models of consumption can be used to obtain samples of $P_L,Q_L$: $P_L^{(i)},Q_L^{(i)}$ for $i \in \{1,\dots,N\}$, and compute the samples $\abs{V}^{(i)}$ using \eqref{eq:PFlin}. Then, using the approaches in \cite{picallo2017twostepSE},\cite{bolognani2017fast}, the few real-time measurements available can be used to update the samples $\abs{V}^{(i)}$. With these samples, we can solve the stochastic optimization problem \eqref{eq:OPFstoch} using our scenario approach \eqref{eq:probfeassuprem}. Note that from \eqref{eq:PFlinincr} the constraints for each scenario $i$ can be represented as
\begin{equation}\label{eq:PFsample}
V_{\min}\textbf{1} \leq \abs{V}^{(i)} + \text{diag(\hspace{-0.05cm}\abs{V_0}\hspace{-0.1cm})}^{-1} (Z_p \Delta P_G + Z_q\Delta Q_G) \leq V_{\max}\textbf{1}.
\end{equation}
For the optimization problem with constraints \eqref{eq:PFsample}, it is not possible to know beforehand the number of support constraints. However, it is possible to identify the most limiting scenarios for each element $(\cdot)_l$ of these vector constraints:
\begin{equation}\label{eq:PFsampleelem}\arraycolsep=1pt\begin{array}{rl}
V_{\min} - \min_i \abs{V}_l^{(i)} & \leq \big( \text{diag(\hspace{-0.05cm}\abs{V_0}\hspace{-0.1cm})}^{-1} (Z_p \Delta P_G + Z_q\Delta Q_G) \big)_l \\
V_{\max} -\max_i \abs{V}_l^{(i)} & \geq \big( \text{diag(\hspace{-0.05cm}\abs{V_0}\hspace{-0.1cm})}^{-1} (Z_p \Delta P_G + Z_q\Delta Q_G) \big)_l.
\end{array}
\end{equation}
Using \eqref{eq:PFsampleelem}, it becomes simple to identify which scenarios are support constraints and which ones will be violated if removed. Hence, we can apply our scenario approach \eqref{eq:probfeassuprem} using the following steps:
\begin{enumerate}
\item Identify the necessary number of scenarios $\tilde{N}$ using \eqref{eq:probfeas} with $\epsilon=0.05$ and $\beta=10^{-3}$. Initialize the set of scenarios $\tilde{\mathcal{N}}$ and the set of removed constraints $I=\emptyset$.
\item Solve the scenario optimization problem \eqref{eq:sceremprob} using the set of scenarios $\tilde{\mathcal{N}}$ and $\mathcal{A}(\tilde{\mathcal{N}})=I$.
\item Observe the number of support constraints $k$. Compute the number of scenarios to remove $R$ using \eqref{eq:probfeassuprem}, such that $\epsilon(k,R)\leq \epsilon<\epsilon(k,R+1)$. Identify the indices $\{i_1,\dots,i_R\}$ of $R$ support constraint looking at \eqref{eq:PFsampleelem}, and add them to $I$: $I = I \cup \{i_1,\dots,i_R\}$. 
\item Repeat step 2). If the new solution has the same number of support constraints $k$ as the previous solution, and all constraints in $I$ are violated, finish here. If not, go back to step 3).
\end{enumerate}

\begin{figure}
\centering
\includegraphics[width=7.5cm,height=6cm]{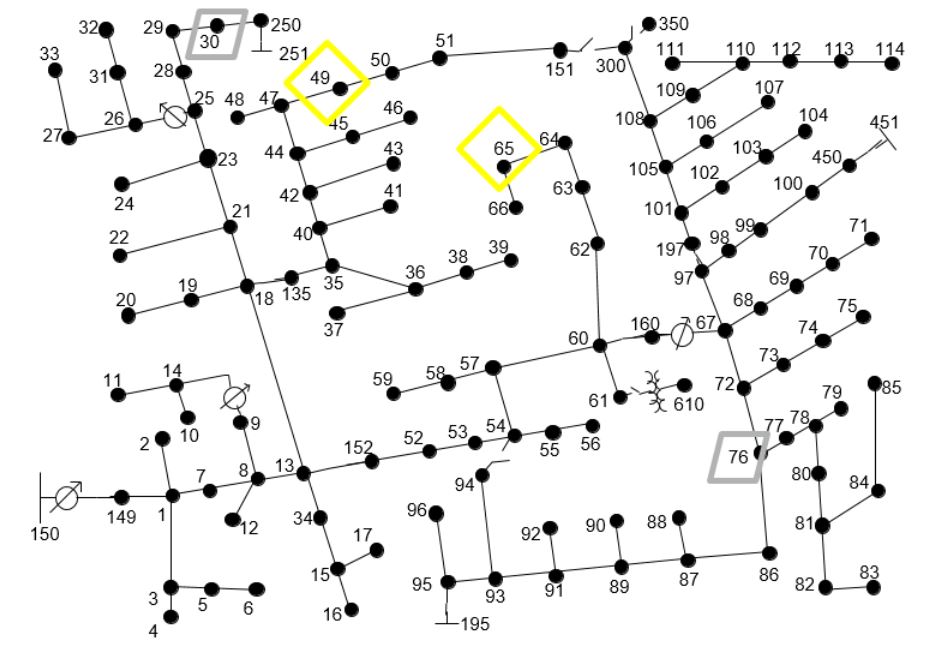}    
\caption{123-bus test feeder with location of distributed generation: a yellow rhombus for solar, a grey parallelogram for wind. The network image has been taken from \cite{testfeeder}.}
\label{fig:123bus}
\end{figure}

\afterpage{
\begin{figure}[t]
\centering
\includegraphics[width=8.5cm]{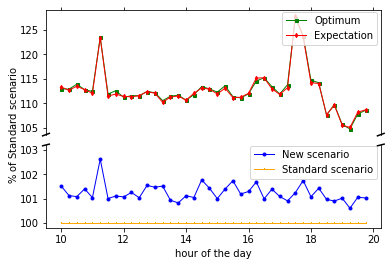}
\caption{Percentage of the total generation of each approach with respect to the \textit{Standard scenario}.}
\label{fig:sceobjective}
\end{figure}

\setlength{\tabcolsep}{3pt}
\begin{table}[t]
\centering
\begin{tabular}{|l|l|l|l|l|}
\hline
Case & Optimum & Expectation & 
Standard scenario & New scenario \\ \hline \hline
$\max$ & $1.050 \pm 0$ & $1.057 \pm 0.003$ & $1.040 \pm 0.0015$ & $1.041 \pm 0.0014$ \hspace{-0.2cm}\\ \hline
$\min$ & $0.950 \pm 0$ & $0.948 \pm 0.003$ & $0.950 \pm 0.0004$ & $0.950 \pm 0.0003$ \\ \hline
\end{tabular}

\caption{Mean $\pm$ standard deviation during the time period from 10am to 8pm, of the $\max$ and $\min$ voltage magnitude $\abs{V}_\text{new}$ across all nodes in the grid: $\text{mean}_t(\max_l (\abs{V}_{\text{new},t})_l)$, and similarly for $\min$ and standard deviation.
}\label{tab:resultsint}
\end{table}
}

We have tested this new scenario approach \eqref{eq:probfeassuprem} in the benchmark distribution grid 123-bus test feeder \cite{testfeeder}, see Figure \ref{fig:123bus}. As in \cite{picallo2018stochOPF}, we have allocated distributed energy sources in some nodes, see Figure \ref{fig:123bus}; and a real-time current measurement at the branch connecting the feeder to the main grid, as in \cite{bolognani2017fast}. Load samples are constructed by aggregating load data of several households \cite{pedersen2015disc}, and generation availability is constructed using real solar irradiation and wind speed data \cite{picallo2018stochOPF}. We choose the standard voltage limits $V_{\min}=0.95, V_{\max}=1.05$ \cite{picallo2018stochOPF}. We consider four approaches:
\begin{itemize}
\item \textit{Optimum}: using the true values of the voltage magnitudes as if they were known.
\item \textit{Expectation}: using the estimated expected values and disregarding the uncertainty.
\item \textit{Standard scenario}: using the standard scenario approach described in \eqref{eq:sceproblem},\eqref{eq:probfeas}.
\item \textit{New scenario}: using the new scenario approach in \eqref{eq:probfeassuprem}.
\end{itemize}
and use them in a simulation to solve the stochastic optimal power flow \eqref{eq:OPFstoch} once every 15 minutes in the interval from 10am to 8pm of a day.

 As it can be observed in Figure \eqref{fig:sceobjective}, both \textit{Optimum} and \textit{Expectation} allow to inject around $10-15\%$ more renewable energy than the scenario approaches \textit{Standard scenario} and \textit{New scenario}. Despite that, in Table \ref{tab:resultsint} it can be observed how the extreme values of \textit{Expectation} are continuously violating the voltage limits: the maximum and minimum voltage magnitude values for all nodes have means beyond the limits $0.95, 1.05$, even with a gap larger than two standard deviations in the case of the maximum. This is a consequence of disregarding the uncertainty in the data. On the other hand, both \textit{Standard scenario} and \textit{New scenario} satisfy the constraints. Moreover, note in Figure \ref{fig:sceobjective} that the \textit{New scenario} injects around $1-2\%$ more energy (even up to $3\%$) than the \textit{Standard scenario}, since some scenarios are removed. In this case, this improvement could imply millions of profit if taking into account the huge amount of distribution grids across the world. An intuitive explanation of what happens, is that out of many possible load conditions/scenarios in an electrical grid, there will be only a few that are actually limiting the current solution, the support constraints. Therefore, if removing some scenarios, the solution will perform better, but still satisfy the probability guarantee on the risk.

\section{Conclusions}\label{sec:conclusions}

In this work we have presented an extension of the scenario approach that allows to sieve out unnecessary constraints, when observing that a problem is not fully supported and thus too conservative. Our methodology determines the number of constraints to remove given the number of support constraints. This allows to improve the performance of the solution, while preserving the probabilistic guarantees. We have shown the relevancy of the approach using an illustrative example; then we have tested its effectiveness in a real-world power system application for solving the optimal power flow problem under uncertainty.
As future work, it would be interesting to analyze if tighter bounds can be derived.

\appendices

\section{Proof of Theorem \ref{thm:remsup}}\label{sec:appproofthm}

This proof combines elements from the proofs in \cite{campi2011sampling,campi2018wait}. We start by defining the problem, where samples with indices in the set $I=\{i_1,\dots,i_R\}$ are removed, with $|I|=R$:
\begin{equation}\label{eq:scefixI}
x_{I}^* = \arg\hspace{-0.3cm}\min_{x \in \mathcal{X} \subseteq \mathbb{R}^d} c^T x \text{ s.t. } x \in \bigcap_{i \in \tilde{\mathcal{N}} \setminus I } \mathcal{X}_{\delta^{(i)}}.
\end{equation}
Consider also the set of samples violating the $I$ constraints:
\begin{equation*}
\Delta_I^{\tilde{N}} = \{\delta^{\tilde{N}} \in \Delta^{\tilde{N}} : x_I^*(\delta^{\tilde{N}}) \notin \mathcal{X}_{\delta^{(i)}} \; \forall i \in I\} \subseteq \Delta^{\tilde{N}}.
\end{equation*}  

Since the solution $x_{\tilde{N},R}^*$ almost surely violates $R$ constraints, we know that, for some $I$, $x_{\tilde{N},R}^*=x_I^*$, and $S_{\tilde{N},R}=S_{I}$, where $S_{I}$ is the set of support constraints of $x_I^*$. Therefore, we have
\begin{equation*}
\begin{array}{l}
\{\delta^{\tilde{N}} \in \Delta^{\tilde{N}} : V(x_{\tilde{N},R}^*(\delta^{\tilde{N}}))>\epsilon(|S_{\tilde{N},R}^*|,R)\} \\[0.1cm]
\subseteq \bigcup_I 
\{\delta^{\tilde{N}} \in \Delta_I^{\tilde{N}} : V(x_{I}^*(\delta^{\tilde{N}}))>\epsilon(|S_{I}^*|,R)\},
\end{array}
\end{equation*}
and thus
\begin{equation}\label{eq:PFsuprem1}
\begin{array}{l}
P(V(x_{\tilde{N},R}^*)>\epsilon(|S_{\tilde{N},R}^*|,R)) \\[0.2cm]  
\leq \sum_I P(\delta^{\tilde{N}} \in \Delta_I^{\tilde{N}} \wedge V(x_{I}^*)>\epsilon(|S_{I}^*|,R)) \\[0.2cm]
\stackrel{a)}{=} \binom{\tilde{N}}{R} P(\delta^{\tilde{N}} \in \Delta_I^{\tilde{N}} \wedge V(x_{I}^*)>\epsilon(|S_{I}^*|,R)) \\[0.2cm] 
 = \binom{\tilde{N}}{R} P(V(x_{I}^*)>\epsilon(|S_{I}^*|,R) \wedge x_I^* \notin \mathcal{X}_{\delta^{(i)}} \; \forall i \in I),
\end{array}
\end{equation}
where the equality $a)$ is due to all possible combinations of indices of the removed constraints $I$. Since the scenario sampling process is i.i.d, the probability is the same for any set of indices $I$ with same size.

Now we can look into this last probability term: $P(V(x_{I}^*)>\epsilon(|S_{I}^*|,R) \wedge x_I^*(\delta^{\tilde{N}}) \notin \mathcal{X}_{\delta^{(i)}} \; \forall i \in I)$. Without loss of generality, we fix the set of removed constraints $I=\{N+1,\dots,N+R\}$, so that the remaining scenarios are $\tilde{\mathcal{N}}\setminus I = \mathcal{N}$. As in Section \ref{sec:supsce}, we consider having a number $k$ of support constraints $S_I^*$, i.e., $|S_I^*|=k$, of problem \eqref{eq:scefixI}:

\begin{equation}\label{eq:PFsuprem2}\arraycolsep=1pt\begin{array}{l}
P(V(x_I^*)>\epsilon(|S_{I}^*|,R) \wedge x_{I}^*  \notin \mathcal{X}_{\delta^{(i)}}\; \forall i \in I) \\[0.2cm]
= \hspace{-0.05cm} P(\bigcup\limits_{k=0}^d \hspace{-0.1cm} \{V(x_I^*) > \epsilon(k,R) 
\wedge x_{I}^* \notin \mathcal{X}_{\delta^{(i)}} \forall i \in I \wedge |S_I^*| = k \} ) \\[0.2cm]
\stackrel{b)}{=} 
\sum\limits_{k=0}^d P(V(x_I^*)>\epsilon(k,R) 
\wedge x_{I}^*  \notin \mathcal{X}_{\delta^{(i)}} \forall i \in I \wedge |S_I^*|=k) \\[0.2cm]
\stackrel{c)}{=} 
\sum_{k=0}^d \binom{N}{k} P( V(x_I^*)>\epsilon(k,R)  
\wedge x_I^* \notin \mathcal{X}_{\delta^{(i)}} \; \forall i \in I  \\
\hspace{2.3cm} \wedge \; |S_I^*|=k \wedge S_I^*=\{1,\dots,k\}) ,
\end{array}
\end{equation}
where equality $b)$ is due to having disjoint sets since the $|S_I^*|$ can only be a single value; and $c)$ is due to all possible combination of indices of the support constraints $S_I^*$ out of the remaining samples $N$ after removing $R$ from $\tilde{N}$. Again, since the scenario sampling process is i.i.d, the probability is the same for any set of indices $S_I^*$ of the same size.

Let $x_k^*$ be the solution using only the first $k$ scenarios:
\begin{equation*}\arraycolsep=1pt\begin{array}{rl}
x_{k}^* = \arg\min_{x \in \mathcal{X}} c^T x \text{ s.t. } &  x \in \bigcap_{i=1,\dots,k} \mathcal{X}_{\delta^{(i)}}. 
\end{array}
\end{equation*}

With Assumption \ref{ass:nondeg}, the solution $x_I^*$ is non-degenerate, i.e., if $S_I^*=\{1,\dots,k\}$, then $x_I^*=x_k^*$ with probability $1$, and we have
\begin{equation}\label{eq:PFsuprem3}\arraycolsep=1pt
\begin{array}{l}
P( V(x_I^*)>\epsilon(k,R) \wedge x_I^* \notin \mathcal{X}_{\delta^{(i)}} \; \forall i \in I \wedge |S_I^*|=k  \\
\hspace{0.45cm} \wedge \; S_I^*=\{1,\dots,k\}) \\[0.2cm]
\stackrel{d)}{=} P( V(x_k^*)>\epsilon(k,R) \wedge x_k^* \notin \mathcal{X}_{\delta^{(i)}} \; \forall i \in I  \wedge |S_k^*|=k \\
\hspace{0.8cm} \wedge \; x_k^* \in \bigcap_{i=k+1,\dots,N} \mathcal{X}_{\delta^{(i)}} ) \\[0.2cm]
= \int_{(\epsilon(k,R),1]} P(x_k^* \notin \mathcal{X}_{\delta^{(i)}} \; \forall i \in I \wedge x_k^* \in \bigcap_{i=k+1,\dots,N} \mathcal{X}_{\delta^{(i)}} \hspace{-0.1cm} \hspace{0.1cm} \big\rvert  \\
\hspace{2.3cm} V(x_k^*)=\upsilon \wedge |S_k^*|=k) dF_k(\upsilon) \\[0.2cm]
\stackrel{e)}{=} \int_{(\epsilon(k,R),1]} (1-\upsilon)^{N-k}\upsilon^R dF_k(\upsilon),
\end{array}
\end{equation}
where $F_k(\epsilon) = P(V(x_k^*)\leq \epsilon \wedge |S_k^*|=k)$ is the $k$-th cumulative density function, i.e., the probability that $x_k^*$ is fully-supported and has violation probability less than $\epsilon$. The equality $d)$ can be proven similarly as in \cite[5.1 Proof of Theorem 1, proof A=B]{campi2018wait}. The equality $e)$ comes from the fact that for a fixed violation probability value $\upsilon$, the probability of satisfying the $N-k$ constraints with indices $\{k+1,\dots,N\}$ is $(1-\upsilon)^{N-k}$; and the probability of violating the $R$ constraints with indices $I$ is $\upsilon^R$.

Putting all together we have:
\begin{equation}\label{eq:PFsuprem4}
\begin{array}{l}
P(V(x_{\tilde{N},R}^*)>\epsilon(k,R)) \\[0.2cm] 
\stackrel{\eqref{eq:PFsuprem1}}{\leq} \binom{N+R}{R} P(V(x_{I}^*)>\epsilon(k,R) \wedge x_I^* \notin \mathcal{X}_{\delta^{(i)}} \; \forall i \in I) \\[0.2cm] 
\stackrel{\eqref{eq:PFsuprem2}\eqref{eq:PFsuprem3}}{=} \binom{N+R}{R} \sum_{k=0}^d \binom{N}{k} \int_{(\epsilon(k,R),1]} (1-\upsilon)^{N-k}\upsilon^R dF_k(\upsilon).
\end{array}
\end{equation}

Now we use additional information about $F_k(\epsilon)$. Similarly as in \cite{campi2018wait}, we can derive some constraints for any generic number of scenarios $m\geq 0$ instead of $N$:
\begin{equation*}\arraycolsep=1pt
\begin{array}{rl}
1 & \geq P(V(x_{I}^*) \geq  0 \wedge x_I^* \notin \mathcal{X}_{\delta^{(i)}} \; \forall i \in I) \\[0.2cm] 
& \stackrel{\eqref{eq:PFsuprem2}\eqref{eq:PFsuprem3}}{=}
\sum_{k=0}^{\min(d,m)} \binom{m}{k} \int_{0}^1 (1-\upsilon)^{m-k}\upsilon^R dF_k(\upsilon),
\end{array}
\end{equation*}
where the term $\min(d,m)$ reflects that for a number of scenarios $m \leq d$, there can only be $m$ support constraints.

So we can derive an upper bound $\gamma$ on $P(V(x_N^*)> \epsilon(s_N^*))$: $P(V(x_N^*)> \epsilon(s_N^*)) \leq \gamma$, by optimizing over all possible $F_k(\cdot) \in \mathcal{C} \; \forall k$, where $\mathcal{C}$ is the positive cone of generalized distribution functions:
\begin{equation}\label{eq:primalgamma}\arraycolsep=1pt\begin{array}{rl}
\gamma = & \sup_{F_k \in \mathcal{C}} \binom{N+R}{R} \hspace{-0.1cm} \sum\limits_{k=0}^d \hspace{-0.1cm} \binom{N}{k} \hspace{-0.1cm} \int_{(\epsilon(k,R),1]} (1-\upsilon)^{N-k} \upsilon^R dF_k(\upsilon) \\[0.2cm]
& \text{s.t. } \hspace{-0.3cm} \sum\limits_{k=0}^{\min(d,m)} \hspace{-0.3cm}\binom{m}{k} \int_{0}^1 (1-\upsilon)^{m-k} \upsilon^R dF_k(\upsilon) \leq 1, \; \forall m \geq 0.
\end{array}
\end{equation}


By truncating $m\leq M$, where $M\geq d$, the number of constraints are reduced. Then we get a new problem with less constraints, whose optimal value $\gamma_M$ satisfies $\gamma \leq \gamma_M$. Now we consider the dual problem of $\gamma_M$\cite{anderson1987linear}:
\begin{equation}\label{eq:dualgammaM}\arraycolsep=1pt
\begin{array}{rl}
\bar{\gamma}_M = & \inf_{ \lambda_m \geq 0} \sum_{m=0}^M \lambda_m \\[0.2cm]
& \text{s.t. } \binom{N+R}{R} \binom{N}{k} (1-\upsilon)^{N-k} \upsilon^R 1_{(\epsilon(k,R),1]}(\upsilon) \\[0.2cm]
& \hspace{0.5cm} \leq \sum_{m=k}^M \lambda_m \binom{m}{k} (1-\upsilon)^{m-k} \upsilon^R\\[0.2cm]
& \hspace{0.5cm} \forall \upsilon \in [0,1], \; \forall k = 0,1,\dots,d
\end{array}
\end{equation}
where $1_{(\epsilon(k,R),1]}(\upsilon)$ is the indicator function. The term $\upsilon^R$ can be canceled on both sides of the constraint, since for $\upsilon=0$ the constraints hold for any $\lambda_m$. By weak duality we have $\gamma \leq \gamma_M \leq \bar{\gamma}_M$. This can be verified for any feasible point $F_k$ of \eqref{eq:primalgamma} and $\lambda_m$ of \eqref{eq:dualgammaM}:

\begin{equation}\label{eq:dualgap}\arraycolsep=1pt
\begin{array}{l}
\binom{N+R}{R} \sum_{k=0}^d \binom{N}{k} \int_{(\epsilon(k,R),1]} (1-\upsilon)^{N-k} \upsilon^R dF_k(\upsilon) \\[0.2cm]
= 
\sum_{k=0}^d \int_0^1 \binom{N+R}{R} \binom{N}{k} (1-\upsilon)^{N-k} \upsilon^R 1_{(\epsilon(k,R),1]}(\upsilon) dF_k(\upsilon) \\[0.2cm]
\stackrel{\eqref{eq:dualgammaM}}{\leq} 
\sum_{k=0}^d \int_0^1 \sum_{m=k}^M \lambda_m \binom{m}{k} (1-\upsilon)^{m-k} \upsilon^R dF_k(\upsilon) \\[0.2cm]
=
\sum_{m=0}^M \lambda_m \sum_{k=0}^{\min(d,m)} \binom{m}{k} \int_0^1  (1-\upsilon)^{m-k} \upsilon^R dF_k(\upsilon) \\[0.2cm]
\stackrel{\eqref{eq:primalgamma}}{\leq} 
\sum\limits_{m=0}^M \lambda_m.
\end{array}
\end{equation}

Then, for any feasible point $\lambda_m \geq 0$ of \eqref{eq:dualgammaM}, we have $\gamma \leq \sum_{m=0}^M \lambda_m$. Let us consider the case $M=N$, and the candidate solution $\lambda_m = \frac{\beta}{N+1} \; \forall m$. To ensure that this is a feasible point of \eqref{eq:dualgammaM}, we need to find the smallest $\epsilon(k,R)$ for every $k \in \{0,\dots,d\}$ such that for all $\upsilon \in [0,1]$
\begin{equation}\begin{array}{c}
\binom{N+R}{R} \binom{N}{k}(1-\upsilon)^{N-k} 1_{(\epsilon(k,R),1]}(\upsilon) \\[0.2cm] 
\leq \frac{\beta}{N+1}\sum_{m=k}^N \binom{m}{k} (1-\upsilon)^{m-k}, 
\end{array}
\end{equation} 
where this equation corresponds to plugging the candidate solution into the constraint in \eqref{eq:dualgammaM}. Now if \eqref{eq:probfeassuprem} has exactly one solution in $\epsilon(k,R) \in (0,1)$, with those $\epsilon(k,R)$ we get
\begin{equation*}
P(V(x_{\tilde{N},R}^*)> \epsilon(|S_{\tilde{N},R}^*|,R) ) \leq \sum_{m=0}^N \frac{\beta}{N+1} - 0 = \beta.
\end{equation*}

The proof that \eqref{eq:probfeassuprem} has exactly one solution in $\epsilon \in (0,1)$, is analogous to the one in \cite[5.3 Proof of Theorem 2]{campi2018wait}. After adding the factor $\binom{N+R}{R}$ the steps of the proof still hold.


\newpage

\bibliographystyle{IEEEtran}
\bibliography{ifacconf}



%

\end{document}